\documentclass[floatfix,showkeys,twocolumn,showpacs,preprintnumbers]{revtex4}
\usepackage{graphicx}
\usepackage{amssymb}

\begin{document}

\title{An agent-based approach to food web assembly}

\author{Craig R. Powell}
\email{craig@powell.name}
\affiliation{Theoretical Physics Group, School of Physics and
  Astronomy, University of Manchester, Manchester, UK, M13 9PL}

\date{\today}

\begin{abstract}
An agent-based model of population dynamics is presented.  The model
has as its expected behaviour the population dynamics of the
equation-based Webworld model, within which large communities of
species can be grown on evolutionary time scales.  Such communities
can be used in the agent-based model without disrupting the food web
structure, and hence a unified model of evolutionary time and
individual-based dynamics can be realised.  Individuals encounter
potential prey, and optimal foraging strategies are arrived at through
natural selection.
\end{abstract}



\maketitle

\section{Introduction}\label{Sec:Introduction}
The basic processes which shape ecosystem functions occur at the level
of individuals \citep{gri05}, whether as direct or indirect
competition between members of the same species, competition between
species, or host/pathogen interactions \citep{bel06,con92} by which
the fitness of an individual might be decreased by the proximity of
other members of the same species.  Where more sophisticated behaviour
exists, mutual benefit may be obtained by co-operation between members
of the same species \citep{ste02}.  The influence of each of these
effects on ecosystem function is, however, perceived only on the time
scale of community assembly.  Species invasions demonstrate how
interactions shape the community on relatively short time scales
\citep{bre08}, and Green Mountain \citep{wil04} exemplifies how an
whole ecosystem can be assembled from disparate parts on a relatively
short time scale.  The features of individual species, though, must
have been arrived at through evolution in the context of a community
of species, and hence only by considering the consequences of
individual interactions on evolutionary time scales can a fuller
understanding of the underlying principles of natural ecosystems be
achieved.  Such questions cannot be readily answered by empirical
study, though microcosm experiments \citep{blo08,sri04} may be able to
address some degree of evolutionary adaptation of species to their
community.  Proper understanding of the mechanisms behind processes
such as behaviour and community assembly and evolution implies the
ability to construct models, numerical or biological, which
demonstrate the sufficiency and necessity of the theories used.  The
importance of heterogeneity at the level of individuals in
understanding ecosystem function is shown by individual- or
agent-based approaches to modelling observed communities
\citep{wys99}.  Likewise, to demonstrate the principles behind
ecosystem evolution requires models able to include these
individual-level effects on evolutionary time scales.

Ecosystem models typically use the simplification of representing
communities as a food web, in which complex interactions are reduced
to feeding relations.  Such models have achieved remarkable success at
reproducing basic observations of real ecology \citep{got06}.  It is
therefore prudent to seek to improve these models by the inclusion of
more realistic interactions, rather than attempting to expand very
detailed models \citep{hol05,reu05,vla04} to include
evolutionary processes.  The Webworld model, first introduced by
\citet{cal98} and modified to include more plausible population
dynamics by \citet{dro01}, has achieved some success in creating food
webs of species either co-evolved in an unchanging abiotic environment
\citep{dro04,qui05b,lug08a}, or constructed through immigration from
such a community \citep{islands}.  By introducing a model whose
expected behaviour is equivalent, but which incorporates stochastic
population dynamics, \citet{stochastic} were able to include within
the context of Webworld a model appropriate for studying population
dynamics on a smaller spatial and temporal scale, the effects of which
need to be considered when applying the earlier model.  It is the aim
of this paper to add a further level of detail to the growing family
of Webworld models, in which individuals can be considered as discrete
and distinguishable entities.  By maintaining equivalence with the
population-level model, it is possible to transfer a model community
between levels of detail so that, in effect, one model can perform
both evolutionary time scales and detailed interactions.  Not only is
this prohibited by computational expense in a single model, but it is
difficult to understand the relation between different scales.  By
manipulation of the different aspects of the Webworld model, insight
can be gained into the relation between phenomena at different scales.

While it may not be possible to model arbitrarily detailed interactions
at the population level, it should be possible to maintain a `ladder'
of models such that at each level consistency can be found with either
the broader or more detailed model.  In doing so, understanding can be
propagated between the two extremes of the model, and concepts of
individual behaviour can hence be translated into ecosystem
consequences.  Once it is demonstrated that an agent-based model
consistent with the insights gained in a population-level model has
been achieved, a number of options are open for future investigation.
These include the introduction of spatial heterogeneity, either as an
explicit effect or by community diversification \citep{rad04}, or of
interactions between individuals beyond those reflected in a food web,
such as mutualism, Janzen-Connell effects \citep{pet08}, or the
effects of direct competition \citep{pic17}

In section~\ref{Sec:model} a summary of the existing Webworld model is
given, followed by the essential features of the new, agent-based
model.  A complete description of the agent-based model, based on the
template of \citet{gri06} and suited to readers wishing to
implement the model, is given in the appendix.
Section~\ref{Sec:analysis} presents analytic results demonstrating
agreement between the agent- and equation-based models in simple
circumstances, and the agreement in small communities of species is
demonstrated by numerical results presented in
section~\ref{Sec:simulation}. Additional insight gained by
the use of agent-based techniques is discussed in
section~\ref{Sec:conc}.

\section{Model description}\label{Sec:model}
The Webworld model was introduced by \citet{cal98}, but significant
changes were made to the population dynamics by \citet{dro01}.  More
recent work has been based on the form of the model in the latter
paper, and it is this that is reproduced by the agent-based model
being introduced.  On the longest timescales, Webworld models the
dynamics of species within food webs as evolutionary changes or
immigration events shift the relative steady-state abundances of the
various species. Species in this sense are described by sets of
attributes, whose exact biological interpretation is assumed to be
unimportant.  For each attribute a species possesses, the fixed matrix
$m$ describes whether it provides a benefit or hindrance during
interactions with an individual of predator or prey species.
Specifically, the score $S_{ij}$ which determines the feeding
interaction between species $i$ and $j$ is the sum over all pairs of
attributes of each species, so
\begin{equation}
  S_{ij}=\frac1L\sum_{\alpha,\beta}m_{\alpha\beta},
  \label{Eq:S}
\end{equation}
where the sum runs over $L$ attributes of each $i$ and $j$.  Each
element of matrix $m$ is a random number chosen from a normal
distribution with zero mean and unit variance.  New species can be
introduced by taking a sub-population of an existing species and
changing one attribute.  The resultant mutant species tends to have
scores similar to those of its parent, but repeated mutations
introduce diversity which allows the formation of complex food web
structures.  A second aspect of the dynamics which relies on the
attribute model is inter-specific competition.  It is assumed that
species which share several attributes are, when they feed on the same
prey species, more directly in competition with each other than if
they had no attributes in common.  The equation used for the
inter-specific competition is
\begin{equation}
  \alpha_{ij}=c+\left(1-c\right)q_{ij},
  \label{Eq:alpha}
\end{equation}
where $q_{ij}$ is the proportion of attributes possessed by both
species $i$ and $j$, and the minimum competition has typically been
chosen as $c=0.5$ in previous papers.

On shorter timescales, the population dynamics is described by the the
equation
\begin{equation}
  \dot N_i=\lambda\sum_jg_{ij}N_i-\sum_jg_{ji}N_j-dN_i,
  \label{Eq:popdyn}
\end{equation}
in which the functional response, $g_{ij}$, is the contribution by
each member of species $i$ to the rate at which individuals of species
$j$ are consumed.  The second sum in (\ref{Eq:popdyn}) is therefore
the total rate at which individuals of species $i$ are lost to
predators; the final term corresponds to death from all other causes,
and is assumed to occur at rate $d$ per individual.  The choice of
$d=1$ sets the timescale of the model.  The first term in
(\ref{Eq:popdyn}) is the total rate at which feeding occurs by member
of species $i$, multiplied by the ecological efficiency,
$\lambda=0.1$.  This factor is the mean number of predator offspring
born for each prey individual consumed.  The functional response of
the parent model incorporates details of adaptive foraging, which is
assumed to adapt sufficiently rapidly that the Evolutionarily Stable
Strategy (ESS) is followed at all times.  The functional response is
given by
\begin{equation}
  g_{ij}=\frac{f_{ij}S_{ij}N_j}{bN_j+\sum_k\alpha_{ik}f_{kj}S_{kj}N_k},
  \label{Eq:g}
\end{equation}
where $b=1/200$ and $f_{ij}$ is some foraging effort divided between
prey species such that, for each species, $i$, $\sum_jf_{ij}=1$.  In this
case the ESS was shown by \citet{dro01} to correspond to
\begin{equation}
  f_{ij}=\frac{g_{ij}}{\sum_kg_{ik}},
  \label{Eq:f}
\end{equation}
which can be solved by iterative application of (\ref{Eq:g}) and
(\ref{Eq:f}).

It was shown in \citet{stochastic} that the parent Webworld model
could be reproduced by a reaction scheme in which sated individuals of
species $i$ become hungry at rate $S_{ij}/b$ according to the last
prey they consumed, and in which all individuals are susceptible to
death with an expected rate $d$.  Foraging remained governed by a form
of the functional response, with foraging predators consuming foraging
prey with a reaction rate coefficient
\begin{equation}
  k\!\left(i,j\right) = \left\{\sum_k\alpha_{ik}f_{kj}\frac{S_{kj}}{S_{ij}}N_k\right\}^{-1}.
  \label{Eq:kijSolved}
\end{equation}
The restriction to foraging prey is based on the insight that prey
behaviour is influenced by the presence of predators to reduce risk
\citep{lim90}.  Prey individuals in the present model can gain no
advantage while sated or tired, and hence should seek refuge.

This paper introduces an agent-based model in which the functional
response is reproduced by heterogeneity of the prey population.  Each
agent is a member of a particular species, which dictates the scores
underlying its ability to feed on particular prey.  The ESS of the
parent model, given by (\ref{Eq:g}) and (\ref{Eq:f}), suggests that it
should not be the optimal strategy to simply attack any prey
individual encountered, and each agent is therefore allocated a
strategy, which indicates the probability with which an individual of
any particular species should be attacked.  In the absence of an
analytic form for this strategy, it should still arise by natural
selection if offspring of each agent adopt a slightly mutated
strategy.

The population heterogeneity necessary to reproduce the Webworld
functional response can be introduced if each agent is given a further
set of attributes, drawn from the same set as the species attributes.
To give an intuitive interpretation these are termed
`vulnerabilities'; they are features which are not inherited, but
which differentiate between the individuals of the same species.  If
the species attributes are those which have been selected by
evolution, the vulnerabilities might correspond to the remaining
variability.  The Webworld model can be reproduced if, in order to
successfully feed on a prey individual, at least one attribute must be
shared by that individual and the predator species.  In this way, the
abundance of the predator attributes is diminished in each of its prey
species through selective predation, while being resupplied through
birth.  Inter-specific competition is intrinsically stronger between
predator species if they share several of the same attributes, since
they will then be feeding on substantially overlapping subsets of the
prey population.  A full specification of the model, based on
\citet{gri06}, is given in the appendix.

To reproduce the Webworld model, it is essential that the
vulnerabilities of a prey individual cannot be perceived directly by
its predators.  A predator will therefore `attack' regardless of
whether the prey individual is vulnerable or not, and hence some
proportion of attempted attacks will be successful.  As predation
pressure increases, the proportion of successful attacks will
decrease, leading to the functional response desired.  If an attack is
successful then the predator individual becomes sated and the prey
individual is killed, but in the event of an unsuccessful attack the
prey individual suffers no penalty, while the predator becomes tired
for some period of time.  Recovery of the Webworld functional
response, given below, requires that this time is equal to the time
taken to recover from the sated state, $b/S_{ij}$.  The intuitive
relation between the two time scales depends heavily on the
interpretation of the tired and sated states.  If the sated period is
thought of as a digestion time, tired individuals would begin foraging
again much more quickly, having nothing to digest.  When the rate of
recovery after an unsuccessful attack is very large, the results
become unlike Webworld, since there is no penalty associated with
attacking a prey for which there is heavy competition.  Conversely, if
the satiation period is thought of as recovery from the physical
exertion of the attack, this may not depend on the degree of success
at all.  It is even possible that the time period for an unsuccessful
attack should be longer than for a successful one; if a predator
gives up an attack after time $t$ then on average successful attacks
must take less than this time, while unsuccessful attacks will always
take time $t$.  The consequences of the interpretation for the prey
individual will not be considered in this paper, since the closest
approximation to Webworld occurs when the prey does not suffer
indirect effects of predation.

\section{Analytic results}
\label{Sec:analysis}
The functional response of the agent-based model can be determined
analytically for simple systems, and useful results are obtained by
considering a prey species of constant population, $N_j$, being
predated by a single predator which itself has no alternative prey.
In this case there are five populations of interest, those being the
number of vulnerable and invulnerable prey, and the number of
foraging, sated, and tired predators.  It is useful to be able to
assume that each population is in a steady-state, and hence set all
time derivatives in the analysis to zero.  By choosing an appropriate
rate coefficient for the death of predator individuals, the predator
population can be large or small compared to the prey population, and
hence the assumption of being in a steady-state does not require a
particular ratio of populations.  During the analysis, $d_i$ will
indicate the death rate coefficient for species $i$.  In the parent
model and in the numerical results of this paper, $d_i=d=1$ for all
species.

In the steady state, the number of individuals of prey species $j$
vulnerable to predator $i$ is given by
\begin{equation}
  v_{ij}=\frac{fd_j}{d_j+\left(1-f\right)\omega N_i^\prime}N_j,
  \label{Eq:v}
\end{equation}
where fraction $f$ of new-born individuals are vulnerable to this
predator, and $N_i^\prime$ predators are foraging concurrently.  The
reaction rate coefficient $\omega=1$ is introduced to clarify
dimensionality.  Because predator individuals are only born of sated
predators, the total predator population is given by
\begin{equation}
  N_i=\lambda\frac{S_{ij}}{bd_i}N_i^*,
  \label{Eq:NPred}
\end{equation}
where there are $N_i^*$ sated individuals.  The number of sated
individuals is given by the balance of feeding with death and the
return to foraging, so (\ref{Eq:NPred}) can be written in terms of the
number of foraging predators as
\begin{equation}
  N_i=\lambda\frac{S_{ij}}{S_{ij}-bd_i}\frac{\omega v_{ij}}{d_i}N_i^\prime.
  \label{Eq:NPred2}
\end{equation}
Assuming that sated predators are very likely to return to foraging
before they die, i.e. $S_{ij}\gg bd_i$, the predator population is
given by
\begin{equation}
  N_i=\lambda\frac{f}{1-f}\frac{d_j}{d_i}N_j.
  \label{Eq:NPred3}
\end{equation}
In the parent model the equivalent result is that $N_i=\lambda N_j$,
with the standard assumption that the death rate is $d$ for all
species.  Thus, the agent-based model is closest to the parent model
for $f=\frac12$.  Given the usual Webworld condition that species have
$L=10$ attributes chosen from $K=500$ possibilities, this occurs most
precisely for 33 vulnerability attributes, since
\begin{equation}
  \left|f-\frac12\right|=\left|\frac{\left(K-L\right)!}{K!}\frac{\left(K-V\right)!}{\left(K-L-V\right)!}-\frac12\right|,
\end{equation}
the `error' in $f$, is minimized for $V=33$.  Using $f=\frac12$, the
rate of change of the predator population can be written as
\begin{equation}
  \dot N_i=\lambda\frac{S_{ij}N_j}{2bN_j+\frac{S_{ij}N_i}{d}+\frac{2S_{ij}}{\omega}}N_i-dN_i.
  \label{Eq:dNPred}
\end{equation}
For large populations, the third term in the denominator is
negligible, and only two differences remain between this functional
response and that used in the parent model, shown in (\ref{Eq:g}).
The first is that there remains a factor of 2 discrepancy in the first
term of the denominator, which relates to high prey abundances.  This
is directly related to the fact that at most half of the prey
individuals are vulnerable to predation in the agent-based model.  The
second difference is the appearance of $d=1$ in the second term of the
denominator.  Since $S_{ij}/b$ is interpreted as a reaction rate, this
is required for dimensional consistency.  One complication of this
analysis is that the effective population of the prey species is the
number of concurrently foraging individuals.  While only a small
fraction of the population are expected to be sated at the same time,
a rather large proportion may be tired.  Thus, in the agent-based
model, the effective prey population may be significantly different
from the total population.  This effect can be removed by denying
tired individuals any immunity to predation, but the numerical
results in this paper examine the case of sated and tired individuals
being `invulnerable'.

In order to remove the remaining need to consider two-body processes,
the foraging process can be replaced by explicit modelling of a small
patch of space.  Individuals move between vertices in a Cartesian
lattice with a characteristic interval, and on arrival at a new site
interact with each individual present in a random order.  Three types
of interaction are possible; in the first case, neither individual
wishes to attack the other, so the subsequent potential interactions
are considered in turn.  Secondly, the new individual may choose to
attack a resident, in which case it must become either sated or tired.
Because interaction with sated and tired individuals is prohibited,
subsequent potential interactions are ignored.  The third possibility
is that the resident individual attacks the newcomer.  If successful,
the newcomer is consumed, and clearly cannot participate in further
interactions.  Otherwise, the attacker becomes tired, but the
subsequent potential interactions of the newcomer are considered in
turn.  If the newcomer does not become sated or tired due to any
interaction, it remains on the new site for the characteristic time,
and may interact with any individuals entering the site.  In this
context, it is more appropriate to consider foraging strategy as the
probability of attacking an individual of a particular prey species
encountered, in which case the strategy of each individual is a value
between zero and one for each prey species.

A useful quantity to consider in order to determine optimal behaviour
is the utility of each individual.  Utility in an ecological context
can be equated to the future lifetime reproductive success of an
individual, and in the case of asexual reproduction it can, for
simplicity, be assumed that each offspring produced corresponds to a
utility gain to the parent equal to its own utility at birth.  If the
probability of dying in a given time interval is not dependent on age,
it follows that the utility depends only on circumstances, and an
individual can take actions to maximize its utility even without
knowing the absolute value.  Writing as $U_i$ the utility of an
individual of species $i$ when not in an encounter, it follows that
the utility of attacking an encountered individual of species $j$,
$U_{ij}$, is
\begin{equation}
  U_{ij}=\frac{S_{ij}}{S_{ij}+bd}\left\{1+\frac{\lambda v_{ij}}{N_j}\right\}U_i,
\end{equation}
where $S_{ij}/\left(S_{ij}+bd\right)$ is the probability of surviving
to start foraging again after the attack, at which time the individual
again has utility $U_i$, and reproduces with probability $\lambda$ if it
encountered one of the $v_{ij}$ vulnerable individuals.  Assuming that
each individual of species $i$ acts in the same way, and attacks
members of species $j$ with probability $p_{ij}$, deductions can be
made about $p_{ij}$.  If $U_{ij}<U_i$, utility is lost by attacking,
so $p_{ij}=0$.  If no related predators feed on species $j$ then half
the individuals of that species will be vulnerable, and the condition
that $p_{ij}=0$ is
\begin{equation}
  S_{ij}<\frac{2bd}\lambda.
\end{equation}
In this case, the chance of dying before starting to forage again
outweighs the potential gain through reproduction.
If the prey species is, instead, strongly favoured, $p_{ij}=1$.  In
this case, assuming that $v_{ij}$ is constant,
\begin{equation}
  \frac{v_{ij}}{N_j}=\frac{d}{2d+\omega N_i^\prime}.
  \label{Eq:vij}
\end{equation}
By considering all the various sub-populations of species $i$ to have
constant population, it can be shown that
\begin{equation}
  N_i^\prime=\left\{1+b\omega\sum_j\frac{p_{ij}N_j}{S_{ij}+bd}\right\}^{-1}N_i.
  \label{Eq:NiPrime}
\end{equation}
It follows that the prey species should always be attacked subject to
the condition that
\begin{equation}
  S_{ij}>\frac{2bd}\lambda+\frac{bN_i}\lambda\left\{\frac1\omega+b\sum_j\frac{p_{ij}N_j}{S_{ij}+bd}\right\}^{-1}.
\end{equation}
For all cases in which $0<p_{ij}<1$, the utility of attacking a
presenting prey individual is equal to the utility of not doing so.
However, this is only the case because the fraction of vulnerable
individuals is maintained by the attack probability.  For each attack
not carried out, the proportion of vulnerable individuals rises, and
it becomes optimal to attack.  Conversely, too many attacks deplete
the abundance of vulnerable prey, and it is optimal to seek other prey
species.

\section{Simulation results}
\label{Sec:simulation}
\begin{figure}[tb]
  \includegraphics[width=0.25\textwidth]{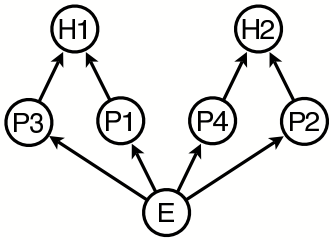}
  \caption{
    The food web used in simulated time-series.  The food web consists
    of two herbivores each feeding predominantly on pairs of plants.
    However, for each pair of species a feeding link is possible in
    one direction, so more completely connected webs can exist for
    sub-optimal foraging.  The species designations are given in
    ascending order of population in the parent model.  The
    environment species, E, has either a fixed population or a fixed
    birth rate.
  }
  \label{Fig:IslandA}
\end{figure}

\begin{figure}[tb]
  \includegraphics[width=0.45\textwidth]{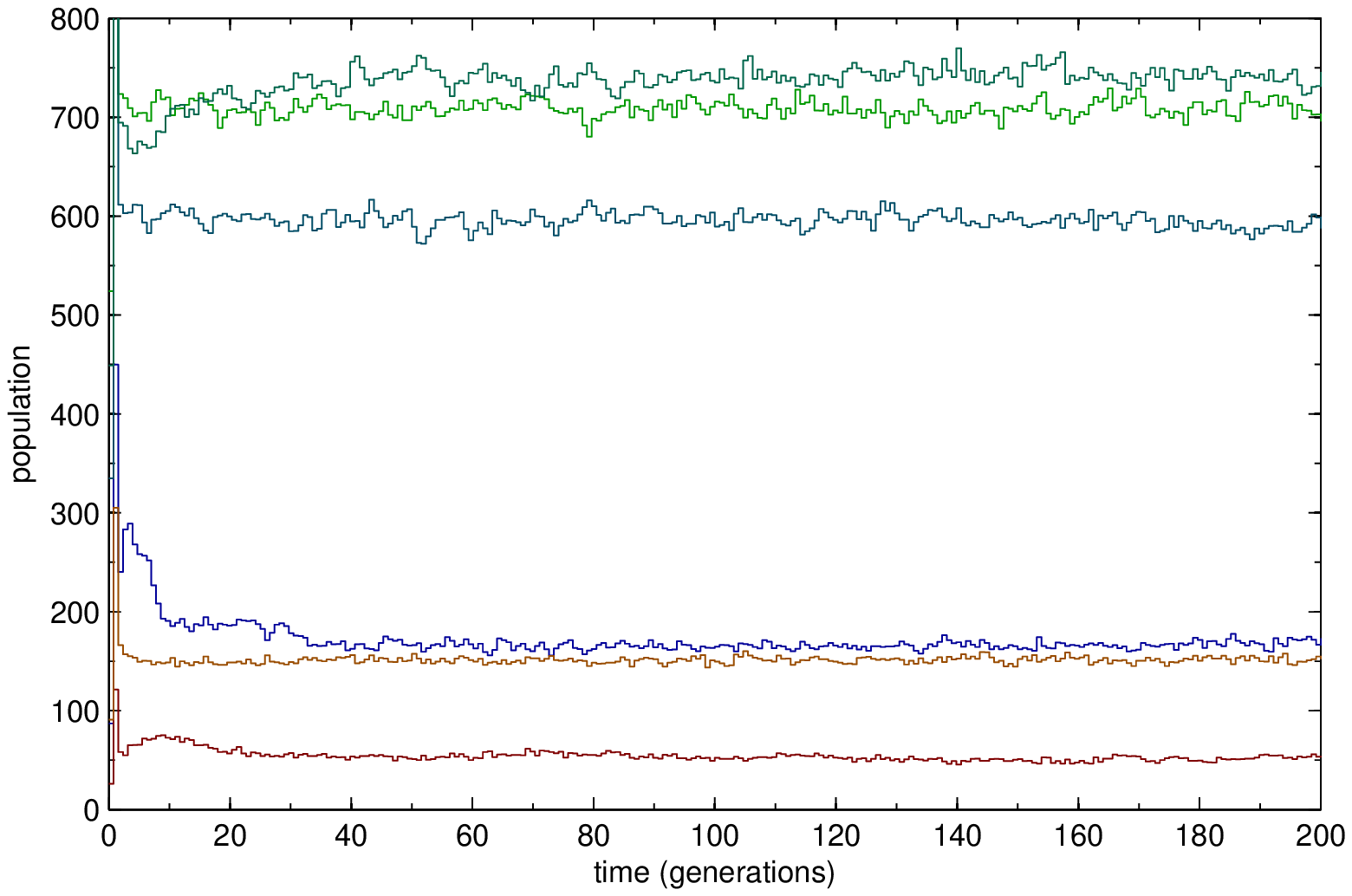}
  \caption{
    The ensemble-average time-series of 20 simulation runs, sampled at
    approximately unit intervals.  In order of ascending population at
    time 200 the species are; H1, H2, P1, P2, P4, and P3.
  }
  \label{Fig:pureEnsemble}
\end{figure}

Since it is not trivial to identify the steady-state distribution of
vulnerability traits in the population of each species, the
simulations examined in this paper each start from a uniform
distribution.  The food web shown in Fig.~\ref{Fig:IslandA}, a small
community grown in the equation-based model, was used to provide a
community of species with established feeding links.
Individuals of the resource species, E, are added at a
constant rate, $\rho$, and are subject to `death' at rate $d$ in the
same way as all other species.  It was experimentally determined that,
for a $100\times100$ lattice with periodic boundary conditions, an
interval between agent movements of $\tau=9174^{-1}$ was needed to
match the unit expectation time in the parent model of two individuals
meeting.  The closest match to the steady-state strategy of the parent
model was built by allocating individuals of each species to pure
foraging strategies according to the ESS, $f_{ij}$ as calculated from
(\ref{Eq:f}).  The strategy of new-born individuals was copied from
their parent with mutation, such that the probability of attacking
each prey species was adjusted by 5\%, then rescaled to make the
greatest probability $1.0$.  Fraction $\mu=0.01$ of new-borns were
assigned an additional prey species chosen at random from the ESS,
which they attack with probability $0.5$ if it was not
already assigned a non-zero probability.  From these initial
conditions, Figure~\ref{Fig:pureEnsemble} shows that after
approximately twenty generations, the population of each species has
settled to an approximately steady value.  The ratio of these
populations are not quite the same as for the parent model, but it is
sufficiently remarkable that the model is able to support food webs
built in the parent model without substantial reconfiguration, given
the gap between the competition models.

\begin{figure}[tb]
  \centering
  \includegraphics[width=0.45\textwidth]{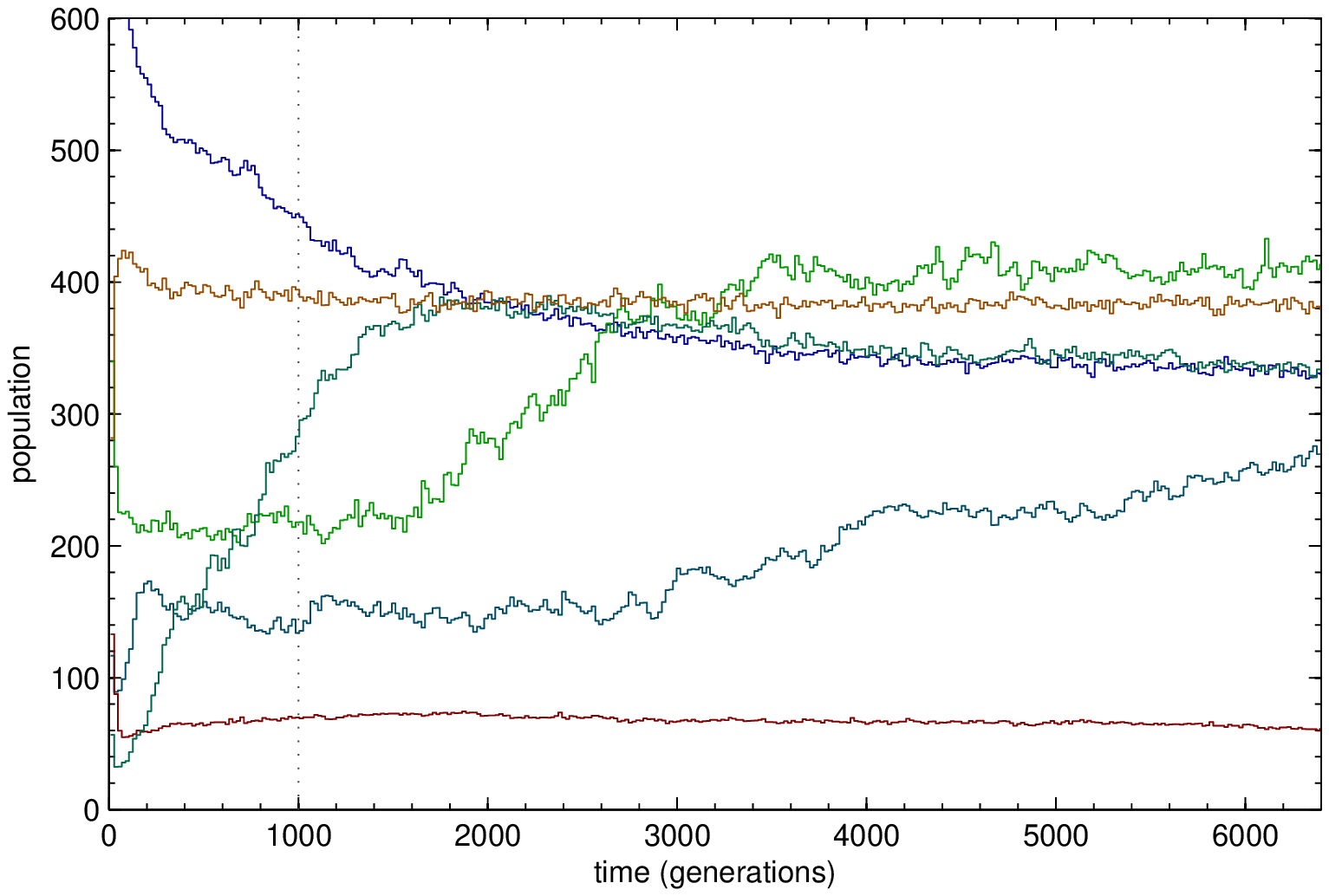}
  \caption{
    The ensemble-average time-series of 20 simulation runs, binned to
    intervals of 20 generations.  In order of ascending population at
    time 1000 (marked) the species are; H1, P2, P4, P3, H2, and P1.
  }
  \label{Fig:glutEnsemble}
\end{figure}

\begin{figure}[tb]
  \centering
  \includegraphics[width=0.45\textwidth]{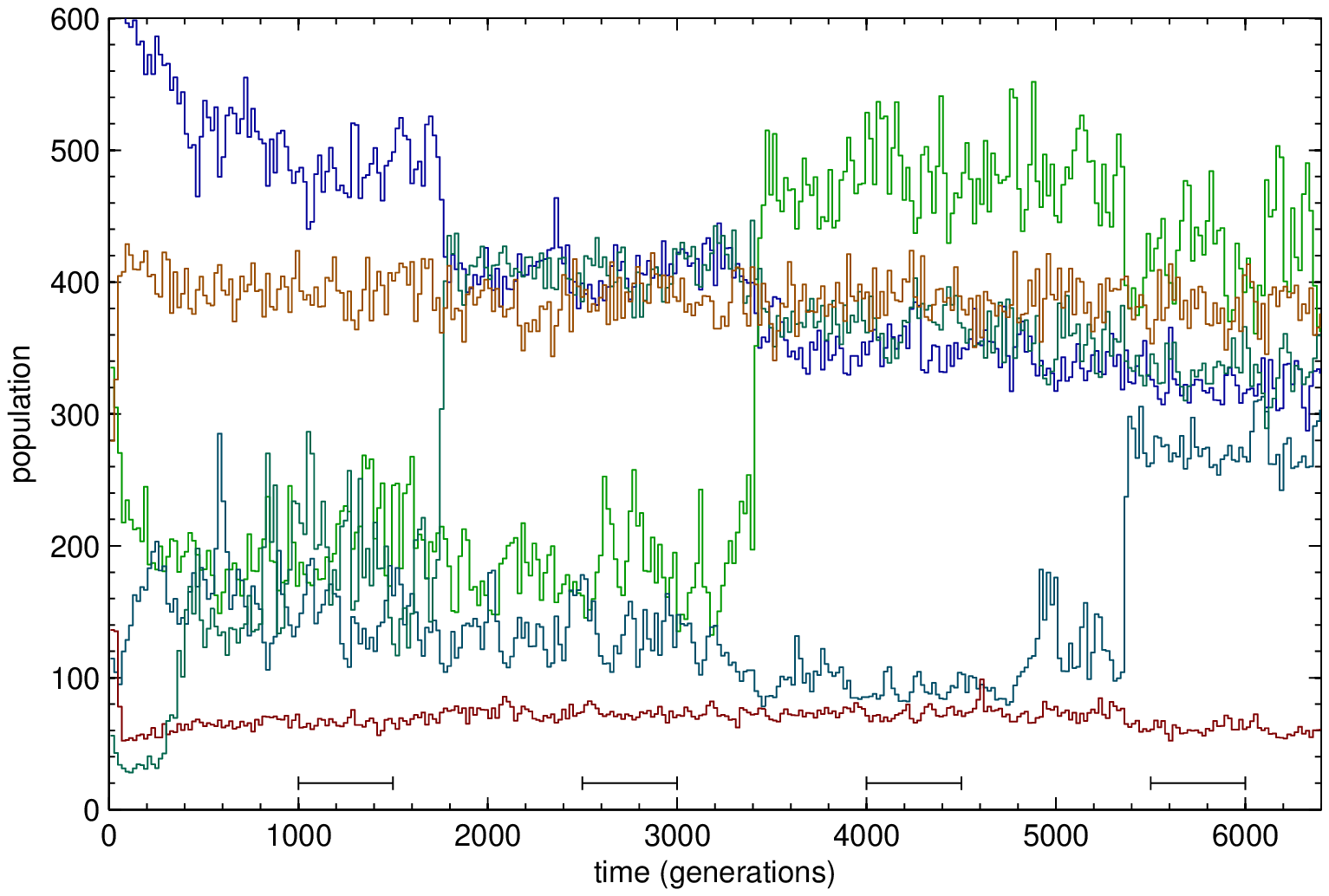}
  \caption{
    A sample time-series, binned to intervals of 20 generations.  The
    feeding strategy of the agents was determined within each of the
    four marked intervals.  Species P1 and H2 initially have the
    highest and second-highest populations respectively, while H1 has
    the lowest population after time 500.  The three transitions to
    larger populations are made, in chronological order, by species
    P3, P4, and P2.
  }
  \label{Fig:glutSample}
\end{figure}

For comparison, the agent-based model was run from an initial
condition in which all prey corresponding to positive $S_{ij}$ are
attacked.  In this case, for each pair of species $i,j\ne i$ either
$S_{ij}$ or $S_{ji}$ is positive, and hence one species will
definitely attack the other when encountered.  Where $S_{ij}$ is very
small this can be very far from the optimal strategy, to the extent
that no feeding link exists in the parent model.  The ensemble-average
time series shown in Figure~\ref{Fig:glutEnsemble} covers a much longer
period of time than does Figure~\ref{Fig:pureEnsemble}, yet even after
this period the system has not reached a steady state.  Examination of
the single time series shown in Figure~\ref{Fig:glutSample}
illustrates the underlying cause of the gradually changing populations
seen in Figure~\ref{Fig:glutEnsemble}.  In any given simulation run,
there exist states in which the population of each species is not
changing systematically with time to any great degree, but which are
interrupted by rapid transitions to a different configuration of
populations.  The change takes place over a few generations, and is
caused by the rapid propagation of a superior foraging strategy
through one species.

The four periods marked in Fig.~\ref{Fig:glutSample} clearly differ in
the relative populations of the six species and, apart from stochastic
effects, do not appear to contain systematic changes in those
populations.  To understand the cause of the differences, the
effective food web was calculated for each period by counting the
number of occasions on which each species consumed each possible prey.
The essential difference between successive periods is the loss of a
single feeding strategy by one species.  By the time of the first
marked period, only three strategies had been lost from the
fully-connected food web of the initial condition; species P3 was no
longer fed on by either species P2 or P4, accounting for its increase
in abundance after an early low, and P4 had also ceased feeding on P2.
By the second period, P1 had also stopped feeding on P3, and it
stopped feeding on P4 by the third period.  In each case the predator
population was decreased by adopting a strategy for which individual
fitness was improved.  By the final period, H1 had stopped feeding on
P2, but two very significant differences from the parent configuration
remain.  In the first place, `plant' P1 still fed on P2 rather than
becoming truly basal.  Secondly, and with great significance to the
relative populations, H2 still fed upon the environment directly.

Allowed sufficient time, the food web configuration present in the
ensemble of Figure~\ref{Fig:pureEnsemble} might be recovered, but the
timescale is extremely long.  An increase in the resource availability
by a factor of two appears to slow the transitions by approximately a
factor of four, but reducing the resources by a similar factor results
in almost all cases in the extinction of P3 during its early
under-abundant phase.  It has therefore not been possible to test for
the complete convergence of the ensembles shown in
Figures~\ref{Fig:pureEnsemble} and \ref{Fig:glutEnsemble}, and
consequently it has not been demonstrated that the agent-based model
will converge on the same configuration as the parent model.

\section{Conclusions}
\label{Sec:conc}
An agent-based approach is a natural level of detail at which to
examine a number of phenomena important to the formation of food web
structures in related models.  In particular, evolutionary mechanisms
operate at this level, and predictions regarding the population-level
behaviour of an evolving system need to be tested by comparison with
the predictions of simulations of individuals, especially where other
experimental confirmation is not possible.  A surprising outcome of the
simulations of the agent-based model presented in this paper is that
the approach to the evolutionary stable strategy assumed by earlier,
population-level models is inhibited by increased populations to such
a degree that the ESS may never be relevant to a food web continually
perturbed by invasions, extrinsic fluctuations, and spatial
inhomogeneity.  The computational demands of agent-based simulation,
as compared to equation-based models of population dynamics, make the
more detailed model inappropriate for examining evolutionary
time-scales, but an investigation of the assembly of immigrant
communities under agent-based dynamics is feasible.  An essential
aspect of future investigation is to relate the agent- and the
equation-based models not merely in terms of mean behaviour, but in
understanding and predicting the deviations of the former model from
that mean.

Two related aspects of the approach to agent-based modelling used are
of particular value.  On the one hand, the process of developing an
agent-based model provides insight into implicit assumptions about the
mechanisms being used.  Where population-level equations have been
chosen to reproduce observed phenomena this can lead to understanding
of the particular interactions which are necessary to reproduce the
observations.  On the other hand, much insight from behavioural
biologists and ecologists can be incorporated into a model only at the
level of interactions between individuals, since the consequences of
including these effects are not know a priori.  The development of a
generic food web model into which these effects can be incorporated is
therefore of great potential value, especially as the model is
already known to be able to reproduce a number of ecological results
relating to food web function and species populations.

\section*{Acknowledgements}
I would like to thank Alan McKane and Richard Boland for comments on
this paper, and Emma Norling and Bruce Edmonds for on-going
discussions about agent-based modelling.  I would like to thank EPSRC
(UK) for funding under grant number GR/T11784.


\appendix
\section{Model Description}
This section provides a specification for the agent-based Webworld
model based on the template of \citet{gri06}.  Further details of the
model at the population level can be found in \citet{cal98} and
\citet{dro01}.

\subsection{Purpose}
The purpose of the model is to reproduce the Webworld model of food
web assembly in an agent-based formulation.  The derivation of the
model gives insight into the mechanisms implicitly underlying the
earlier model, and the use of an agent-based framework lays the
foundations for future abstract models to incorporate more realistic
ecological detail.

\subsection{State variables}
The model comprises species, identical in description to those used in
earlier, non-agent-based papers, individuals, and a spatial lattice.
Individuals are characterised by the state variables: species,
vulnerabilities, strategy, birth date, date of next action, a
direction, and a three-valued discrete variable recording whether the
individual is foraging, tired, or sated.  Global variables are given
in Table~\ref{Tab:global}.
\begin{table}
  \begin{tabular}{c|c|c}
    Symbol & Description & Value \\
    \hline
    $b$       & Time scale for `digestion' & 0.005 \\
    $d^{-1}$  & Maximum life span & 1.0 \\
              & of individual \\
    $\lambda$ & Ecological efficiency/ & 0.1 \\
              & probability of reproduction \\
    $\mu$     & Probability of acquiring & 0.01 \\
              & new prey \\
    $\rho$    & Rate of introduction of & 50\,000 \\
              & resource individuals \\
    $\tau$    & Time interval for movement & $9174^{-1}$ \\
  \end{tabular}
  \caption{
    Global constants of the agent-based Webworld model.  $\rho$
    is chosen to determine the final population size and/or number of
    species.
  }
  \label{Tab:global}
\end{table}

The species is a set of $L=10$ attributes chosen from $K=500$
possibilities.  For each pair of species $i,j$, these determine the
score of $i$ when feeding on $j$, $S_{ij}=-S_{ji}$, calculated using
(\ref{Eq:S}).  Negative values of $S_{ij}$ are never used.  Species
are evolved by mutation and selection as described by \citet{dro01}.
A constraint on the set of attributes is that the $L$ attributes must
all be different, and that species must differ in at least one
attribute.

The vulnerabilities are a set of $V=33$ attributes chosen from the
same set as the species attributes.  These are chosen at birth from a
uniform distribution, with the condition that the $V$ attributes are
all different.

The strategy of an agent is an associative array of species and
probabilities.  The array must contain at least one species with
associated probability 1, and does not contain any species, $j$, for
which $S_{ij}\le0$, where $i$ is the species of the agent.  For each
species not represented in the array, the probability is assumed to be
zero.  Conceptually, the offspring of an agent adopt the same set of
prey species with slightly altered probabilities, and therefore cannot
normally feed on prey species which were unknown to their parent.

The birth date and date of next action are used to determine the life
history of the individual as specified in the next section.

The lattice associates each individual with one lattice point, and
represents a two-dimensional Cartesian space with periodic boundaries
to model a well-mixed spatial distribution.

Individuals of the special `resource' species are introduced at rate
$\rho$.  Individuals of this species can be attacked in the same
manner as a normal species, but cannot themselves attack.  They are
assumed to always be in the `foraging' state.
\begin{table}
  \begin{tabular}{c|c}
    Species & Attributes \\
    \hline
    E  &  10,  33,  60,  78, 114, 260, 342, 346, 391, 428 \\
    P1 &  81, 204, 217, 240, 328, 335, 357, 368, 388, 467 \\
    P2 &   1,  43, 156, 165, 210, 220, 250, 320, 368, 481 \\
    P3 & 127, 130, 183, 193, 204, 210, 225, 240, 481, 467 \\
    P4 &  43,  80, 193, 217, 320, 390, 400, 425, 446, 470 \\
    H1 &   5,  31, 224, 332, 367, 374, 450, 459, 463, 495 \\
    H2 &  73, 211, 252, 294, 297, 330, 336, 370, 384, 466 \\
  \end{tabular}
  \caption{
    Attributes of the co-evolved species in the small community used
    for the simulation results of this paper.
  }
  \label{Tab:IslandAAttributes}
\end{table}

\subsection{Process overview and scheduling}
The model proceeds by discrete events, and has an internal date
variable to represent the most recent event.  Each agent acts either
on its internal date-of-next-action or time $d^{-1}$ after its birth,
whichever the sooner.  In the latter case the agent simply dies, and
is removed from the model.  The model identifies the agent with the
earliest date, choosing at random between agents that happen to have
the same date, typically due to the limits of numerical precision.
Before the selected agent acts, new resources are added to the system
according to a Poisson distribution with expectation value $\rho\delta
t$, where $\delta t$ is the difference between the agent's date of
action and the date of the most recent event.  Like `real'
individuals, resources are removed from the model time $d^{-1}$ after
they are added.  The action that the agent takes depends on its state.
If the agent is tired, it enters the foraging state, and schedules a
date $\tau$ later on which to move.  An agent does the same if it is
sated, except that in this case it reproduces with probability
$\lambda$.  The creation of the offspring is detailed below.  If the
agent is neither tired nor sated, it must be foraging, and the event
is one of movement.

Movement occurs on the spatial lattice.  The value of $\tau$ given in
Table~\ref{Tab:global} corresponds to a square lattice of
$100\times100$ points, with periodic boundaries.  To promote mixing,
agents move in the same direction as their previous movement with
probability $\frac12$, or move to the left or right of that direction
with probability $\frac14$ in each case.  Other movement schemes may
be preferred but have not been investigated.  On the arrival of agent
X at the new lattice point, all foraging agents present are placed in
a queue in random order.  Agents which are not foraging cannot predate
on agent X, and are assumed to be concealed such that they cannot be
predated themselves.  For each agent in the queue in turn, agent X and
the queued agent are each given the opportunity to attack one another.
Because $S_{ij}=-S_{ji}$ for any species $i$ and $j$, mutual predation
and cannibalism is forbidden in Webworld, and hence for any pair of
individuals at most one will attack.  If X still exists and is still
in the foraging state, the interaction with the next agent in the
queue is investigated.  If X remains in the foraging state after all
possible interactions have been considered, it schedules a movement
action after time $\tau$.
\begin{table}
  \begin{tabular}{c|cccccc}
    & \multicolumn{6}{c}{Prey} \\
    Predator & E & P1 & P2 & P3 & P4 & H1 \\
    \hline
    P1 & 5.599 && 1.924 & 0.7194 & 0.9849 \\
    P2 & 5.586 &&& 0.2975 \\
    P3 & 5.72 \\
    P4 & 5.885 && 0.1906 & 0.4078 \\
    H1 & & 3.096 & 0.5075 & 4.829 & 0.4491 \\
    H2 & 0.1252 & 1.54 & 4.781 & 0.582 & 2.793 & 0.6788 \\
  \end{tabular}
  \caption{
    Scores of the small community used for the simulation results of
    this paper.  Non-positive scores are omitted for simplicity.
  }
  \label{Tab:IslandAScores}
\end{table}

When agent Y is given the chance to attack agent Z, it examines its
strategy to determine the result, based on the species of Z.  If Y
attacks, it discovers whether or not Z has a vulnerability to Y's
species.  If so, the attack is successful, in which case Z is
destroyed and Y enters the sated state.  If Z is not vulnerable to Y's
species, Y enters the tired state, but Z is unaltered.  In particular,
Z does not cease foraging.  In either case, Y is no longer foraging,
and reschedules its next action to occur after time $b/S_{ij}$, where
$i$ is the species of Y and $j$ is the species of Z.

Offspring are added to the system as copies of their (single) parent
in the same location.  Their birth date is set, fixing the latest date
at which they might die, and they schedule a movement action for time
$\tau$ after their birth.  The vulnerabilities of the offspring are
generated as a random selection of the $K$ species attributes, without
repetition.  For each species in the parent's strategy, the
probability of attacking that species is modified $\pm0.05$, then
scaled such that the maximum probability is 1.  If any `probability'
has become less than or equal to zero, the corresponding species is
removed from the strategy.  With probability $\mu$, the ESS of the
species is calculated using (\ref{Eq:g}) and (\ref{Eq:f}), and one prey species selected using the ESS as
a probability distribution.  If the offspring's strategy does not
contain that prey, it is added with probability of attack equal to
0.5.

\subsection{Design concepts}
\emph{Emergence:} Although there exists an expected behaviour of each
species according to the parent model, the strategy in fact allows the
emergence of a food web configuration given an arbitrary set of
species.  When evolution or immigration of new species occurs, the
balance of introduction and extinction causes an emergent food web
configuration.
\begin{table}
  \begin{tabular}{c|ccccccc}
    Species & E & P1 & P2 & P3 & P4 & H1 & H2 \\
    \hline
    Population & 20000 & 87 & 335 & 449 & 524 & 26 & 91 \\
  \end{tabular}
  \caption{
    Initial populations used in the agent-based simulations.  These
    reflect the steady-state populations of the parent model, scaled
    by a factor of 20 to reduce stochastic extinctions.
  }
  \label{Tab:IslandAPopulation}
\end{table}

\emph{Adaptation:} On short time scales, the strategy of agents is
subject to selection.  The strategy of a population of agents should
approximate the ESS of the parent model, which is found explicitly.
On longer time scales, species are introduced as mutants of existing
species in the same manner as in the parent model.  These species
co-evolve to promote adaptation between predators and prey.

\emph{Fitness:} Fitness is implicitly modelled as the time taken for
an individual to reproduce.  Individuals successful in identifying prey
species susceptible to attack, and which correspond to large score
values, will reproduce rapidly and hence increase in abundance.
Individuals are implicitly unfit if they are vulnerable to abundant
predators, but vulnerability traits are not inherited and hence no
selection operates.

\emph{Prediction:} Agents do not possess any degree of reasoning, and
do not even adjust their strategy according to their life history.
Natural selection adjusts the mean strategy of each species to suit
the expected success rate of feeding on each prey species.

\emph{Sensing:} Individuals are able to sense only the species to
which other individuals belong.

\emph{Interaction:} All interactions between individuals are of a
predatory nature.

\emph{Stochasticity:} The movement of individuals on the lattice is
stochastic, and the decision to attack prey is stochastic if more than
one prey species is known.  On evolutionary time-scales, the mutations
available to species are stochastic, allowing selection to drive the
evolution of the ecology.

\emph{Collectives:} The only collective entities in the model per se
are species, which dictate the possible interactions between
individuals.

\emph{Observation:} Intended observations are the population of each
species, and the strategy averaged over all individuals of each
species.  The latter is a proxy for the strategy, $f_{ij}$, measured
in the parent model.

\subsection{Initialization}
For the results in this paper, the model was initialized from a small
community of species grown in the parent model.  Using an arbitrary
population multiplier, individuals were placed in the model in
proportion to the species population in the parent model.
Vulnerabilities were assigned at random, although this cannot be the
steady-state configuration since individuals vulnerable to extant
predators are depleted.  For the results most like the parent model,
strategies were initially assigned by dividing agents into groups
according to the ESS.  Subject to rounding, fraction $f_{ij}$ of the
individuals of species $i$ were assigned a `pure' strategy consisting
of attacking species $j$ with probability 1.  Agents, and resources in
an abundance specified as an initial condition, were scattered at
random on the lattice with random direction.

For the simulation runs presented in this paper,
Table~\ref{Tab:IslandAAttributes} details the species present, an
overview of the food web being shown in Fig.~\ref{Fig:IslandA}.  The
scores which lead to the feeding relations and the behaviour of the
agent-based model are given in Table~\ref{Tab:IslandAScores}.  All
simulation runs were initialized with the populations shown in
Table~\ref{Tab:IslandAPopulation}.

\subsection{Input}
Resources are added to the system at a constant rate, allowing
resource influx to balance resource consumption and `death'.  Species
can be added as mutants of existing species by changing one attribute,
or from a community evolved previously, perhaps in the parent model.

\end{document}